\documentclass[prd,aps,amsmath,amssymb,letter]{revtex4}
\usepackage{graphicx}
\usepackage[usenames,dvipsnames]{color}
\usepackage{epsfig,latexsym}
\begin{document}
\title{Percolating Cosmic String loops from evaporating primordial black holes}
\author{Ajit M. Srivastava}
\email{ajit@iopb.res.in}
\affiliation{Institute of Physics, Sachivalaya Marg, 
Bhubaneswar 751005, India}

\begin{abstract}
The Pulsar timing data from NANOGrav Collaboration has regenerated interest
in the possibility of observing stochastic gravitational wave background 
arising from cosmic strings. Standard theory of formation of cosmic strings
is based on a spontaneous symmetry breaking (SSB) phase transition 
in the early universe, with a string network forming via the so called 
Kibble mechanism. This string network rapidly evolves and reaches a scaling 
solution  with a given spectrum of string loops and long strings at any given 
time. This scenario necessarily requires that the entire observable Universe 
goes through the SSB phase transition. This would not be possible, e.g., in
models of low energy inflation, where the reheat temperature is much lower
than the energy scale of cosmic strings.  We point out a very different 
possibility, where a network of even high energy scale cosmic strings 
can form when the temperature of the Universe is much lower. 
We consider local heating of plasma in the early universe by evaporating
primordial black holes (PBHs). It is known that for suitable masses of PBHs,
Hawking radiation of evaporating primordial black holes may
re-heat the surrounding plasma to high temperatures, restoring certain
symmetries {\it locally} which are broken at the ambient temperature of 
the Universe at that stage. Expansion of the hot plasma cools it so that 
the {\it locally restored symmetry} is spontaneously broken again.
If this SSB supports formation of cosmic strings, then string loops
will form in this region around the PBH.  Further, resulting temperature 
gradients lead to large pressure gradients such that plasma will develop 
radial flow with the string loops getting stretched as they get dragged 
by the flow. For a finite density of PBHs of suitable masses, one will get
local hot spots, each one contributing to expanding cosmic string loops.
For suitable PBH density, the loops from different regions may intersect.
If that happens, then intercommutation of strings can lead to percolation,
leading to the possibility of formation of infinite string network, even when
the entire universe never goes through the respective SSB phase transition.
\end{abstract}

\maketitle

There is a renewed interest in cosmic strings in view of the Pulsar timing 
array (PTA) data from NANOGrav Collaboration raising the possibility of 
observing stochastic gravitational wave background arising from cosmic 
strings \cite{pta1} (see, also, ref.\cite{pta2}). With PTA, and upcoming 
Laser Interferometer Space Antenna (LISA) \cite{lisa}, search for cosmic 
strings using their gravitational wave signatures has reached new levels of 
excitement \cite{cosmstrgw}. Formation of cosmic strings, and their possible 
observation, is of great importance as they can provide a direct possible 
window to the physics of ultra-high energy scales. Apart from the standard 
spontaneous symmetry breaking related cosmic strings, these also arise
in superstring theories \cite{superst}, making a compelling case for
the existence of cosmic strings in the Universe.  In the conventional picture, 
strings are produced via the so called Kibble mechanism \cite{kbl} when 
the universe goes through a symmetry breaking phase transition. 
This string network rapidly evolves and reaches a scaling 
solution  with a given spectrum of string loops and long strings at any given 
time. This necessarily requires that the entire observable Universe goes 
through the SSB phase transition. This scenario may not always be possible, for 
example, in models of inflation which have low reheat temperature, or 
in models of TeV scale gravity \cite{tev}.  

Here we propose a very different possibility involving evaporating primordial 
black holes (PBHs).  There have been a host of investigations exploring
the possibility that small black holes can form in the early universe.
(See, ref. \cite{pbh0} for a recent review. Also, see refs. \cite{pbh} for 
early discussions.) Black holes evaporate by
emitting Hawking radiation, and for sufficiently small black hole
masses, they can completely evaporate away during early stages of the
universe. There has been extensive discussion in the literature regarding
the possibility of formation of a hot plasma surrounding primordial
black holes (see \cite{heckler1,heckler2,cline,kapusta}). In these
references the plasma consists of black hole radiation and particles
produced by the interaction of the black hole radiation with itself.
However, for PBH embedded in an ambient thermalized plasma in the
early universe, it is rather straightforward to see that the plasma
near the black hole will be heated up as the black hole radiation
propagates through the ambient plasma and loses energy. A detailed
investigation of this was carried out in \cite{ewb} (see, also \cite{ewb2}) 
utilizing known results about the energy loss of quarks and gluons traversing
a region of quark-gluon plasma \cite{eloss}.  (We mention here that the
results of ref.\cite{eloss} are certainly applicable for high energy partons, 
with energies of several 100 GeV, it is not clear whether they can be 
extrapolated to the case of Hawking radiation with energies almost near the 
GUT scale. Still, the overall picture of PBH heating the local plasma to very 
high temperatures may remain valid even for such large Hawking temperatures).  
It was shown in \cite{ewb} that this energy loss leads to rapid heating of 
the plasma near the black hole. Further, resulting temperature gradients
lead to large pressure gradients such that plasma develops radial flow.
One therefore gets the situation that evaporating primordial black holes
in the early universe lead to radially expanding hot plasma regions

For suitable masses of PBHs,
Hawking radiation of evaporating primordial black holes may
re-heat the surrounding plasma to high temperatures, restoring certain
symmetries {\it locally} which are broken at the ambient temperature of 
the Universe at that stage. Expansion of the hot plasma cools it so that 
the {\it locally restored symmetry} is spontaneously broken again \cite{ewb}.
If this SSB supports formation of cosmic strings, then string loops
will form in this region around the PBH.  Further, resulting temperature 
gradients lead to large pressure gradients such that plasma will develop 
radial flow. String loops can get stretched as they flow with this expanding
plasma.  For a finite density of PBHs of similar masses, one will get
local hot spots, each one contributing to expanding cosmic string loops.
For suitable PBH density, the loops from different regions may intersect.
If that happens, then intercommutation of strings can lead to percolation,
leading to the possibility of formation of infinite string network, even when
the entire universe never goes through the respective SSB phase transition.

 Possibility of formation of primordial black holes, and various 
observational constraints on them are well discussed in the literature
\cite{pbh0,pbh}.  In first order phase transitions primordial black holes 
can be produced by collapsing regions of false vacuum \cite{kodamaetal} or 
due to inhomogeneities formed during bubble wall collisions 
\cite{hawkingetal}. They could also form by large amplitude density
perturbations produced due to fluctuations in the inflaton field 
\cite{carrlidsey}, or by shrinking cosmic string loops
\cite{hawking2}.

 For primordial black holes produced from large amplitude density 
fluctuations produced by the inflation, one might ask  whether CMBR 
data puts any constraints. However, as discussed in ref. \cite{green},
one should note that CMBR data imposes restrictions on the power 
spectrum at {\it large scales} (of order 1 - $10^3$ Mpc), whereas, 
the primordial black holes form due to large amplitude and {\it much
smaller  scale} density fluctuations produced by the inflation. Hence 
they do not affect CMBR results at all but imposes constraints on the 
spectral index and consequently puts restrictions on certain models of 
inflation.  The constraints on primordial black hole formation come from 
other sources like nucleosynthesis, gamma ray background etc. \cite{green}.
However, all these phenomena essentially impose constraints on primordial
black holes with a mass range which is much larger than the ones 
relevant here. We consider primordial black holes with masses at most
few tonnes and such black holes evaporate away much above the QCD scale,
without causing any conflict with current observations.
Our main aim is to illustrate a completely novel possibility of forming
infinite string networks via black holes. Hence, we will simply assume
the required masses and number densities of black holes, without
discussing any specific mechanism which could give rise to the formation 
of such black holes.

Consider the case when the plasma heated by the black hole leads to
local restoration of a gauge symmetry (with an energy scale and associated
phase transition temperature of order $\eta$) which allows for the 
existence of cosmic strings \cite{kbl}.  As the plasma 
flows out radially away from the black hole, its temperature will decrease 
and fall below $T_c (= \eta)$ at some distance $r_\eta$ where the symmetry 
will be broken. Near that region, again, string defects will form either
via the Kibble mechanism, or due to turbulence. (For internal symmetries,
spatial variations of the order parameter could lead to formation of
non-trivial windings when turbulent motion of the plasma folds up
extended spatial regions, essentially compactifying parts of spatial 
regions, i.e. plasma). 

If the friction forces dominate over the
string tension so that strings are  effectively frozen locally in the
plasma then the string loops will be carried out by the radially expanding 
plasma. They will then stretch out to large sizes instead of shrinking
due to tension. Note that cosmic string loops shrink when energy can be 
dissipated in other modes, such as gravitational waves, in particle
emission, or dissipation in background plasma. For strong friction case 
(which is likely in the early stages of hot plasma) strings can expand, 
with the plasma flow increasing the energy of the expanding string loop.

If there is
a uniform density of similar primordial black holes, then each black hole
will emit string loops which will be stretched to large sizes as they
are carried away by convective flow away from each black hole. We now
recall a very crucial property of string-string interaction. It is
well established that string defects when crossing each other, intercommute. 
This always happens as long as strings are not crossing with 
ultra-relativistic velocities \cite{crs}. Thus intercommutation will almost 
always occur here as string velocities are not expected to increase beyond 
the sound velocity. If a string loop emitted by one black hole intersects
the string loop emitted by a nearby black hole, this intercommutation
will lead to formation of a much larger string loop. Iteration of
this process where string loops of different black holes start
intercommuting with each other leads to the remarkable possibility
that these string loops may percolate and lead to formation of an infinite
string network. This possibility is remarkable because such an infinite
string network would normally only arise when the entire universe
undergoes phase transition. 

It is interesting to note here that this picture
is quite the opposite of the standard evolution picture of cosmic string
network which are formed in an overall  SSB phase transition. In that case,
large string loops fold on themselves, and intercommutation at the 
intersecting point leads to the large string loop breaking into smaller
loops, which decay away by emitting radiation (gravitational waves, or
particles). This is the most crucial aspect of the scaling behavior of
the cosmic string network evolution. Completely opposite to this, in our
model, string loops first expand (due to expanding plasma). Intercommutation
of these expanding smaller loops leads to formation of much larger loops, 
eventually forming infinite string network. However, This {\it reverse}
evolution only happens during the period when black hole evaporation
fuels expanding plasma shells. After the black holes evaporate away
(within one Hubble period of the beginning of this process), 
one may expect that the final infinite string network will
be similar to the conventional string network, and may undergo
same evolution, achieving scaling solution eventually. However, there  
is a subtle point here \cite{refr}.  In the conventional scenario, the
strings form due to random variation of the order parameter field 
(for U(1) case, the phase of the scalar field), in
uncorrelated domains leading to non-trivial windings. In contrast, 
in the present scenario, expanding hot plasma produces string
loops (which have zero net winding  for regions outside the loop)
in the background of a relatively cold Universe which initially has
topologically trivial order parameter  configuration (ignoring any
previously present strings). Thus, on global 
scales, order parameter field behaves very differently in the two cases.
It may then be possible that there is some difference in the nature
of the string network in the two cases. We will discuss this issue
further in the following, after discussing the details of the model.

  Let us discuss the conditions for percolation of
strings. A black hole of mass $M_{bh}$ evaporates 
by emitting Hawking radiation with an associated
temperature 

\begin{equation}
T_{bh} = {M_{Pl}^2 \over 8 \pi M_{bh}} 
\end{equation}
Here $M_{Pl} = 1.2 \times 10^{19} GeV$ is the Planck mass. We use natural 
units with $\hbar$ = c = 1. The rate of loss of mass by the
evaporating black hole is given by

\begin{equation}
{dM_{bh} \over dt} = - {\alpha M_{Pl}^4 \over M_{bh}^2} ~.
\end{equation}

Here, $\alpha$ accounts for the scattering of emitted particles by
the curvature and depends on $T_{bh}$. For different values of
$T_{bh}$ values of $\alpha$ have been tabulated in \cite{scat}.
For $T_{bh} = 1,~ 200$, and 10$^{15}$ MeV, the corresponding values of
$\alpha$ are $3.6 \times 10^{-4}$, $2.3 \times 10^{-3}$ and 4.5 $\times
10^{-3}$ respectively. For the range of black hole temperatures for
our model, we will set $\alpha$ to be $3 \times 10^{-3}$.
The lifetime $\tau_{bh}$ of the black hole can be obtained by
integrating Eq.(2). We get

\begin{equation}
\tau_{bh} \simeq 10^2  M_{Pl}^{-4} M_0^3 .
\end{equation}

\noindent where $M_0$ is the initial mass of the black hole.
Eq.(2) implies that very little energy is emitted until time of the order of
$\tau_{bh}$ which is when  most of the energy of the black hole gets emitted.
Thus for a black hole formed early in the Universe, it is reasonable to
assume that the black hole essentially evaporates only when the age of the
Universe is of order $\tau_{bh}$.

Let us assume that black holes formed
in the early Universe have masses so that they evaporate when the
temperature of the Universe is $T_U$.
The age of the Universe $t_U$ when its temperature is $T_U$ is

\begin{equation}
t_U \simeq 0.3 g_*^{-1/2} M_{Pl} T_U^{-2}, 
\end{equation}

where $g_* (\simeq 100$) is the number of relevant degrees of freedom 
\cite{kolb}. By equating $\tau_{bh} 
= t_U$, we can get the mass $M_0$ of the black hole such that its 
evaporation becomes effective when the temperature of the Universe is $T_U$,

\begin{equation}
M_0 =0.07 M_{Pl}^{5/3} T_U^{-2/3} ~.
\end{equation}

Black hole with this mass will have the temperature,

\begin{equation}
T_{bh} = 0.6 M_{Pl}^{1/3} T_U^{2/3}\, .
\end{equation}

For example, with $T_U$ = 1 GeV we get $M_0 = 4 \times 10^{11} M_{Pl}$, $T_{bh}
= 10^6$ GeV and $\tau_{bh} = 5\times 10^{17} GeV^{-1}=
3 \times 10^{-7}$ s.  The picture then is that these black holes emit
particles with energies roughly equal to $T_{bh}$ (= $10^6$ GeV for this
sample case) into the background
plasma which is at a temperature $T_U$ ($\sim$ 1 GeV) to start with.
These $10^6$ GeV particles will scatter with the particles
in the background plasma and will heat it up through their energy loss.
For the black hole masses considered here
only elementary particles will be emitted,
such as quarks, gluons, photons, leptons, etc.
(We mention here that the mass of a primordial black 
hole can also increases due to accretion of background plasma particles.
However, for relevant ranges of black hole masses here, all the dominant
particle species  are ultra relativistic, so only the geometric cross-section
of black hole is relevant which is not very effective \cite{acrt}.
Some slow growth of the mass of the black hole could occur because of
this, and the black hole masses we use here should be taken to be the final
mass of the black hole when its evaporation becomes effective.)

In ref. \cite{ewb}, conditions for the equilibration of 
the Hawking radiation in the ambient plasma were discussed.  Further, 
estimates of plasma velocity were made in \cite{ewb} using
the Euler equation for a relativistic fluid and
it was shown that plasma velocity rapidly becomes relativistic.
This is natural to expect because of sharp temperature gradients
close to the evaporating black hole.
Assuming a steady state situation where the luminosity $L(r)$ is
independent of $r$ and equal to $-dM_{bh}/dt$, we can get
the relation between the temperature $T$ of the expanding plasma shell
and the radius $r$ of the shell (see, ref.\cite{ewb} for details).

\begin{equation}
r=\Biggl[{L \over \gamma^2 (8\pi^3 g_*/45) T^4 v }\Biggr]^{1\over2} \,.
\end{equation}

where $\gamma$ is the Lorentz factor for the plasma velocity.
This expression is valid for distances where bulk plasma flow
dominates over diffusion of particles. As shown in \cite{ewb}, this will
be applicable for the entire range of distances of relevance to our
scenario in the present paper.  We will take the plasma velocity to
be of order sound velocity $ v \simeq 1/\sqrt{3}$. Though it is
important to realize that much larger velocities will be expected
during last {\it explosive} stages of black hole evaporation,
and shocks should develop during those stages. We will not consider 
shocks to keep calculations simple. (Hence, for estimates, we ignore
the Lorentz factor in the above equation.)

 Eq.(6) gives the temperature of the black hole when its evaporation
becomes significant. Within one Hubble time the entire black hole
evaporates away. However, Eqs.(1),(2) show that as the black hole mass
decreases, its temperature, and the rate of energy loss, also increase. 
Thus it is the last stages
of black hole evaporation which will lead to highest temperatures
for the plasma in the nearby region. As our interest is in determining
how far the string loops can be dragged by expanding hot shells,
we need to consider the plasma flow during these last stages of
black hole evaporation. Initial mass $M_0$ of the black hole is determined
by the ambient temperature $T_U$ when black hole evaporation becomes
significant. However, we consider the situation only when its
mass has reduced to a value $M_x$, given by

\begin{equation}
M_x = {M_0 \over x}, ~~~ x > 1
\end{equation}

$M_x$ is determined by the following considerations.
Let us assume that the cosmic strings under consideration correspond
to a symmetry breaking scale $\eta$. (For simplicity we will consider
gauge strings for the estimation of string tension, but ignore the
length scales associated with gauge fields.)
Thus black holes which are relevant to
our scenario must have temperatures larger than $\eta$. Strings
will be produced in expanding shells as they cross the
critical distance $r_\eta$ where temperature drops to a value below the
critical temperature, which we take to be of order $\eta$.
For the self consistency of this picture, we must have the thickness
$\Delta r$ of the plasma shell to be at least of the order of $\eta^{-1}$
which is roughly the thickness of the string. With a spherical shell around 
the black hole, with this thickness, several correlation domains will
then form, leading to string formation via the Kibble mechanism \cite{kbl}.
Thickness of the shell is determined by the life time of the remaining 
black hole with mass $M_x$.  We thus require (using Eq.(3)),

\begin{equation}
100 M_{pl}^{-4} {M_0^3 \over x^3} \ge \eta^{-1}
\end{equation}

Initial temperature $T_x$ of this remaining black hole is given by Eq.(1)
with $M_{bh} = M_x$ (Eq.(8)). We take this temperature to be effectively at
a distance of order of the Schwarzschild radius  $r_x = {2 \over M_{pl}^2}
{M_0 \over x}$. (This is reasonable within an order of magnitude, see
discussion in \cite{ewb}.) Using the temperature distance relationship
from Eq.(7), we can get the temperature of the plasma from this remaining
(last stage) of the black hole as,

\begin{equation}
	T(r) \simeq 0.06  M_{pl} ~ \sqrt{{x \over r M_0}}
\end{equation}

 We mention here that though Eq.(7) (and hence Eq.(10)) was derived in 
ref.\cite{ewb} with considerations of bulk plasma and particle diffusion,
essentially  similar profile is obtained if we take $ T = T(r_x)$ effectively 
at a distance of order of the Schwarzschild radius  $r_x = {2 \over M_{pl}^2} 
{M_0 \over x}$ along with the condition $T^4 r^2 =$ constant. (Which 
essentially means conservation of energy-momentum of the expanding shells). 
We will be considering black hole masses such that $T_x > \eta$.
Temperature will drop to the value $\eta$ at a distance $r_\eta$ which is
obtained by setting $T(r_{\eta}) = \eta$ in the above equation. 
We get,

\begin{equation}
r_\eta = ({0.06 M_{pl} \over \eta})^2 ({x \over M_0})
\end{equation}

String loops will be produced at this distance. We then expect these
string loops to be dragged (and consequently stretched) by expanding
plasma.  At certain distance $r_U$, the temperature will drop to the ambient
value $T_U$ and plasma flow should stop roughly at that distance.
String stretching is not possible beyond distance of this order.
$r_U$, therefore, sets an upper limit on the distance 
$R_{stretch}$ up to which strings can be carried out by plasma flow.
Using Eqs.(5),(10), we can get $r_U$ by setting $T(r_U) = T_U$. we
get,

\begin{equation}
r_U \simeq 0.05 x {M_{pl}^{1/3} \over T_U^{4/3}}
\end{equation}

It should be clear that $r_U$ sets an upper limit on the distance
$R_{stretch}$ up to which strings can be carried out by plasma flow. 
To determine $R_{stretch}$ we need to find the 
condition when the string motion is dominated
by friction forces. Friction forces on strings arise from the scattering
of plasma particles from the string. It has been discussed in the literature
that the dominant contribution to this comes from Aharanov-Bohm scattering
(of particles with appropriate). For simplicity, we
will assume this to be the case. Friction force per unit length on the
cosmic string is then given by \cite{frict},

\begin{equation}
F_{frict} \sim \beta T^3 {v \over \sqrt{1 - v^2}} ~,
\end{equation} 

\noindent where $v$ is the string velocity through the plasma, $T$ is the 
plasma temperature, and $\beta$ is a numerical parameter related to
the number of relevant particle species \cite{frict}. In specific grand 
unified theory models \cite{fracq} it could be of order one,
and we will assume that to be the case here. (See, also, ref. \cite{evrt} 
for discussions of scattering of particles from strings.) We mention 
here that stretching of string due to plasma flow has been discussed by 
Chudnovsky and Vilenkin in ref. \cite{curv} where they considered 
light superconducting cosmic strings getting stretched due to turbulent 
plasma flow in the galactic disc and even in stellar interiors. One  could
also consider formation and stretching of such strings around primordial
black holes. This will require considerations of magnetic fields in the
cosmic plasma etc.

  We consider a simple situation when a string encircling the black hole
is stretched symmetrically by radial flow of the plasma. (Order of 
magnitudes of our estimates should remain same for other geometries of 
strings, e.g. string ends may lie on the boundary of the symmetry
restored region around the black hole)
Such large strings should easily form during string
formation on the surface of the sphere with radius $r_\eta$ near the
black hole.  String tries to collapse due to its tension. 
For a string loop of radius $R$, one can estimate the tension force as
\cite{curv},

\begin{equation}
F_{tension} \sim {\mu \over R} ~=~ {\eta^2 \over R}
\end{equation}

At the formation stage, near $r \simeq r_\eta$, the
string will be at rest w.r.t local plasma frame. In such a situation
$v$ in Eq.(13) will be zero and there will be no friction force on
the string. This will lead to collapse of the string under its tension.
As the string loop starts collapsing, it will develop non-zero velocity
w.r.t. the local plasma frame, and consequently a non-zero friction
force via Eq.(13). As the plasma flow velocity is taken to be the sound
velocity (= $1/\sqrt{3}$) w.r.t. the black hole rest frame, we can deduce
that the friction force on the string will be large as the string
collapses, and will decrease as string collapse decreases. When
string is also at rest w.r.t. the black hole, then the friction force
will be given by Eq.(13) with $v = 1/\sqrt{3}$, and should balance the
string tension force (Eq.(14)). String loop will expand because of
plasma drag as long as $F_{frict}(v=1/\sqrt{3}) > F_{tension}$.
Using $r$ dependence of the temperature from Eq.(10), this condition
gives the upper limit for $R$ as,

\begin{equation}
R < 5 \times 10^{-8} {\beta^2 x^3 M_{pl}^6 \over M_0^3 \eta^4}
	\equiv R_{max}
\end{equation}

 As plasma flow stops beyond a distance of order $r_U$, we conclude that 
plasma flow can stretch string loops to sizes of order $R_{stretch}
\simeq min(R_{max}, r_U)$.  We note that the largest values of 
$R_{max}$, (as well as $r_U$ (Eq.(12)), and $r_\eta$ (Eq.(11)), correspond 
to  the largest value $x_{max}$ of $x$ obtained by taking equality sign in 
Eq.(9), and we will use this. 

We should also consider the effect of
black hole gravity on the string loops. Gravitational force per unit length
on the string due to the black hole of mass $M_{bh}$ can be roughly
estimated as ${G M_{bh} \mu \over R^2}$ where R is the separation of the 
string segment from the black hole. (We are taking string as a simple
gravitating system, as for a string loop. For special geometries, such
as for a straighter string the gravitational force will be different, and
will be typically smaller). We see that,
per unit length, this gravitational force becomes much less than the 
force $f_{tension} \sim {\mu \over R}$ due to string tension when
$R > G M_{bh}$. Thus for any distances much larger than the Schwarzschild
radius of the black hole, gravity of the black hole is sub-dominant
compared to string tension forces. We are considering the situation
when plasma drag forces completely dominate over the string tension forces, 
leading to stretching of string to large distances (where black hole gravity 
becomes even more negligible). Thus black hole gravity is not 
relevant for string dynamics  in our case.

 If black hole number density is such that inter-black hole separation
$d_{bh}$ is smaller than $R_{stretch}$ then loop intersections will
be frequent. As string velocity is not expected to be ultra-relativistic
here (neglecting shocks), this will lead to intercommutation of
strings and hence percolation of strings resulting in an infinite
string network. By assuming that energy density of these black holes
(with mass $M_0$) contributes only a fraction $f$ of the total energy 
density of the universe, we get (with $g_* \simeq 100$),

\begin{equation}
d_{bh} \simeq 0.13 {M_{pl}^{5/9} \over f^{1/3} T_U^{14/9}}
\end{equation}

 Percolation of loops (leading to infinite string network formation)
will happen when $d_{bh} < R_{stretch}$. We can now determine
conditions on various parameters
which satisfy all the constraints for string percolation. 
For given values of the fraction $f$, and string scale $\eta$, the 
condition for percolation ($d_{bh} < R_{stretch}$) shows that
percolation can happen if the Universe temperature $T_U$ lies in the
range $T^{min}_U < T_U < T^{max}_U$, which is determined as follows
(using largest value of $x$ from Eq.(9), as we discussed above).
Note that as $R_{stretch} \simeq min(R_{max}, r_U)$, condition for
percolation requires that $d_{bh}$ should be smaller than both
$r_U$ and $R_{max}$.

If $R_{stretch} = r_U$, then the condition $d_{bh} < r_U$ implies

\begin{equation}
T_U < 3 \times 10^{-2} f^{3/4} M_{pl}^{1/4} \eta^{3/4} \equiv T^{max}_U
\end{equation} 

If $R_{stretch} = R_{max}$, then the condition $d_{bh} < R_{max}$ implies

\begin{equation}
T_U > { 1.6 \times 10^2 \eta^{27/14} \over M_{pl}^{13/14} f^{3/14}}
\equiv T^{min}_U
\end{equation}

\begin{table}[h]
\begin{center}
\caption{Allowed parameters (in GeV units) for percolation of string loops}
\begin{tabular}{||ccccccccc||}
\hline
f &$\eta$&$T_U$&$r_\eta$&$r_U$&$R_{max}$&$d_{bh}$&$M_0/M_{pl}$&$M_x/M_{pl}$\\
\hline
1.0 & $10^{10}$ & $10^5$ & $1.9 \times 10^{-6}$ & $1.9 \times 10^4$ & $7.4 \times 10^2$ & 87.0 & $1.7 \times 10^8$ &  230.0 \\
$10^{-7}$ & $10^{10}$ & $10^6$ & $1.9 \times 10^{-6}$ & 190.0 & 740.0 & {\bf 520.0} & $3.7 \times 10^7$ &  230.0 \\
$10^{-6}$ & $10^{10}$ & $5 \times 10^5$ & $1.9 \times 10^{-6}$ & 740.0 & 740.0 & 710.0 & $5.9 \times 10^7$ &  230.0 \\
1.0 & $10^7$ & 1.0 & 0.19 & $1.9 \times 10^{13}$ & $7.4 \times 10^{11}$ & $5.2 \times 10^9$ & $3.7 \times 10^{11}$ &  $2.3 \times 10^3$ \\
$10^{-3}$ & $10^7$ & 1.0 & 0.19 & $1.9 \times 10^{13}$ & $7.4 \times 10^{11}$ & $5.2 \times 10^{10}$ & $3.7 \times 10^{11}$ &  $2.3 \times 10^3$ \\
$10^{-7}$ & $10^7$ & 10.0 & 0.19 & $1.9 \times 10^{11}$ & $7.4 \times 10^{11}$ & $3.1 \times 10^{10}$ & $8.0 \times 10^{10}$ &  $2.3 \times 10^3$ \\
1.0 & $10^4$ & 1.0 & $1.9 \times 10^4$ & $1.9 \times 10^{12}$ & $7.4 \times 10^{20}$ & $5.2 \times 10^9$ & $3.7 \times 10^{11}$ &  $2.3 \times 10^4$ \\
$10^{-3}$ & $10^4$ & 1.0 & $1.9 \times 10^4$ & $1.9 \times 10^{12}$ & $7.4 \times 10^{20}$ & $5.2 \times 10^{10}$ & $3.7 \times 10^{11}$ &  $2.3 \times 10^4$ \\
$10^{-7}$ & $10^4$ & 1.0 & $1.9 \times 10^4$ & $1.9 \times 10^{12}$ & $7.4 \times 10^{20}$ & $1.1 \times 10^{12}$ & $3.7 \times 10^{11}$ &  $2.3 \times 10^4$ \\
$10^{-10}$ & $10^7$ & 7.0 & 0.19  & $3.8 \times 10^{11}$ & $7.4 \times 10^{11}$ & ${\bf 5.4 \times 10^{11}}$ & $1.0 \times 10^{11}$ &  $2.3 \times 10^3$ \\
$10^{-10}$ & $10^5$ & 0.1 & 410.0  & $4.0 \times 10^{14}$ & $7.4 \times 10^{17}$ & $4.0 \times 10^{14}$ & $1.7 \times 10^{12}$ &  $1.1 \times 10^4$ \\
$10^{-10}$ & $10^5$ & 0.05 & 410.0  & $1.6 \times 10^{15}$ & $7.4 \times 10^{17}$ & $1.2 \times 10^{15}$ & $2.7 \times 10^{12}$ &  $1.1 \times 10^4$ \\
\hline
\end{tabular}
\end{center}
\end{table}

 In Table 1 we have given several sets of values for various parameters 
for which string percolation happens. We see that for all these cases,
$R_{stretch} >> r_\eta$, (where $R_{stretch} \simeq min(R_{max}, r_U)$).
This means that, after getting produced near $r_\eta$, the string loops
are dragged (via friction forces) and stretched by the plasma flow up to 
very large distances. There are several points to be noted from numbers
given in table I. We have included two cases, represented by boldfaced
digits in the column of $d_{bh}$. For $d_{bh} = 520.0$ GeV$^{-1}$ 
and $5.4 \times
10^{11}$ GeV$^{-1}$, string loops will not percolate. This is because
in these cases, though $d_{bh}$ is smaller than $R_{max}$, but it is
larger than $r_U$. As string loops can get stretched only up to a
distance $R_{stretch} \simeq min(R_{max}, r_U)$), in these cases, loops
emanating from different black holes will not intersect, hence no
possibility of percolation. We have also chosen parameters to show
cases such that in some case $r_U > R_{max}$, while in other cases
$R_{max} > r_U$, (both lengths being larger than $d_{bh}$ allowing
for string percolation).

We also see that the percolation of strings is not 
a rare occurrence, but happens for rather generic set of parameter values. 
It is important to see that strings with energy scale $\eta = 10^7$ GeV
can percolate and form an infinite network when the ambient temperature
of the Universe is as low as $T_U = 1$ GeV. Similarly, strings with
$\eta = 10^5$ GeV scale can form with $T_U = 50 $MeV. (Though, for such
low temperatures, one should change value of $g_*$ in Eq.(4) etc.).
Further, even with almost negligible fraction of energy density in 
black holes, $f = 10^{-10}$, string percolation is achieved.

 However, for large values of $\eta$, percolation is achieved only
during very late stages of black hole evaporation when its mass
has become very small (i.e. $x$ becomes very large, with mass of
black hole in this late stage being $M_0/x$), only few 100 times the 
Planck mass.
Eqs.(1),(2) describe black hole evaporation process when quantum gravity
effects have been neglected. It can then be of concern whether these
results can be used for such late stages of black hole evaporation when
its mass is not very much larger than the Planck mass. We, therefore, 
consider smaller values of $\eta$ also, for which much larger black holes
masses can also lead to string percolation. Though, we mention that the
numbers given in table I do not represent any optimized set of values of
various parameters. These are just taken as sample examples, showing
that string percolation in this model can happen for a range of parameter
values. For example, small values of $M_x$ here are due to the fact
that we take maximum allowed value of $x$ from the inequality in Eq.(9).
We do that so that we get largest values of $r_U$ and $R_{max}$ (as
we mentioned above). However, even with smaller values of $x$ string
percolation will be possible, if suitable values of other parameters
(e.g. $T_U$) are chosen. We have also ignored shock formation in the
expanding plasma. During last stages of exploding black hole, shocks
will be expected and they can significantly add to the possibility of
expanding string loops attaining large sizes. A more detailed investigation 
is needed for all these considerations to better optimize these 
parameter values.

 We mention that we have considered here black holes of similar
 masses as an example. Clearly, the basic physics of this model will
 remain applicable for black holes with a range of masses. Though,
 simplest situation for this model will be to have a black hole mass 
 spectrum which is peaked at some specific value, for example,
 if the black holes form in a first order phase transition 
 \cite{hawkingetal}.

We now come back to the issue we mentioned above in the 
beginning relating to possible differences between the final infinite
string network in the present model, and the conventional string network
formed via the Kibble mechanism. First, we clarify that we have not
actually demonstrated that an infinite string network will necessarily form
in this scenario. Even if all requirements for overlapping string loops
(arising from different black holes) are satisfied, including 
intercommutation at string intersection points, one still needs to
check that intersecting loops do percolate to give rise to infinite string
network. Proper answer for this question can be given by a detailed numerical
simulation of this scenario. Even if we accept that an infinite string network
will form in this scenario, there are reasons to expect that there may be
important differences in the nature of string network on large 
scales \cite{refr}. In the conventional scenario of Kibble mechanism,  
the starting point is the existence of uncorrelated domains after an overall
phase transition in the Universe, with random variation of the order 
parameter field (say, the phase of scalar field for U(1) strings) across
different  domains. This, with the use of geodesic rule,  leads to 
non-trivial windings, and hence the formation of strings. We may think
of this conventional scenario as top to bottom model, i.e. "from order
parameter field variations to strings" model. A very important feature of
this scenario is that non-trivial windings are present  at all scales
larger than the correlation domains, and in particular they are 
necessarily present at scales larger than the horizon (Hubble) 
scale \cite{refr}. Indeed, precisely this {\it causality argument} was 
the one which led to the {\it monopole problem} in the Universe, and
which strongly constrains the domain wall models.
The situation, in this respect, is very different in the scenario
we have discussed in this paper. Here, PBHs evaporate in a relatively
cold Universe where the order parameter field has settled to 
a roughly uniform, topologically trivial, configuration (ignoring
any previously present strings). Starting point here is small string
loops forming around evaporating black holes, which have topologically 
trivial winding numbers on scales larger than the loop size. Even during 
the period when these loops expand, due
to plasma expansion, order parameter field remains topologically trivial
on larger scales, and this remains true even if string loops from
different black holes intersect and percolate. This scenario may thus
be taken as bottom to top model, i.e. "from string loops to 
order parameter field variations" model.  Even if an
infinite string network forms in this scenario, it is no longer clear if
actual distribution of order parameter field on large scales, in particular
on Hubble scales, is the same here as in the conventional case. If there
are important differences, it will raise important questions whether the
nature of the infinite string network in the two cases has any crucial
differences. Though, approach to scaling solution does not  retain much 
memory of the initial string network, as it involves robust steps, 
like folding of large loops to self intersection, and small loops 
radiating away. Thus one may expect that such a scaling solution
may be  expected in the present scenario also. Still, it will be very
interesting to explore this issue in detail whether the final string 
networks in the two cases, even in the scaling regime, differ in any 
qualitative manner. It is possible that the cases of gauge strings and
global strings may behave very differently in this respect. One may expect
stronger differences in the two scenarios of string formation for the global
string case, where order parameter field variation plays a more direct role
in the string network evolution. In comparison, for gauge string case,
the string tension is the primary relevant factor for late stages, thus
making it more likely that the string networks in these two scenarios
may have similar late time evolution for the gauge string case.

   We conclude by emphasizing that our results present a possible
scenario for  the formation of an extended (infinite) cosmic string 
network in the Universe without the entire Universe undergoing a
spontaneous symmetry breaking phase transition. Importantly, this
happens when the ambient temperature of the Universe is much below the 
relevant spontaneous symmetry breaking scale.
String loops are generated by evaporating primordial black holes
leading to local symmetry restoration. Expansion of locally heated plasma
cools it, leading to symmetry breaking (again locally), producing cosmic
string loops. Expanding plasma stretches these loops, and intersection
of loops originating from different regions (heated by different PBHs)
can lead to percolation, hence the formation of an extended (infinite)
string network without an overall phase transition.
Though there appear to be some important differences in the distribution
of the order parameter field on large scales in this scenario, as compared
to the conventional Kibble mechanism (as discussed above), the robust
mechanisms responsible for achieving a scaling solution for the string 
network may hold out here also leading to the standard scaling solution. 
It needs to be further explored if, even in the scaling regime, there are
any important differences between the string networks in the two cases.

\vskip .1in

 I would like to thank R. Rangarajan, S. Digal, S. Sengupta, R. Ray, 
B. Layek, and A.P. Mishra  for useful discussions and comments. 


\end{document}